\documentstyle[12pt]{article}
\newcommand{\MeV}{{\rm MeV}}                    
\newcommand{\rme}{{\rm e}}                      
\newcommand{\bold}[1]{\mbox{\boldmath $#1$}}    
\newcommand{\r}{{\bold{r}}}			
\newcommand{\bfC}{{\bold{C}}}			
\newcommand{\z}{{\bold{z}}}                     
\newcommand{\Z}{{\bold{Z}}}			
\newcommand{\zbar}{{\bar{\bold{z}}}}		
\newcommand{\Zbar}{{\bar{\bold{Z}}}}		
\newcommand{\ie}{{\em i.e.}}                    
\newcommand{\beq}{\begin{equation}}
\newcommand{\eeq}{\end{equation}}
\newcommand{\beqar}{\begin{eqnarray}}
\newcommand{\eeqar}{\end{eqnarray}}
\newcommand{\bfig}{\begin{figure}}
\newcommand{\efig}{\end{figure}}
\newcommand{\del}{\partial}                     
\newcommand{\bd}{\begin{itemize}} 
\newcommand{\ed}{\end{itemize}} 
\newcommand{\bc}{\begin{center}}
\newcommand{\ec}{\end{center}}

\newcommand{\be}{\begin{equation}}
\newcommand{\ee}{\end{equation}}
\newcommand{\ba}{\begin{array}}
\newcommand{\ea}{\end{array}}

\newcommand{\Part}{{\cal Z}}       

\newcommand{\SEV}[1]{\prec{#1}\succ} 

\newcommand{\C}{{\bold{C}}}

\newcommand{\ssc}[1]{{\scriptscriptstyle #1}}
\newcommand{\rhoE}{\rho_\ssc{E}}
\newcommand{\comment}[1]{{}}
\newcommand{\figposition}[1]{%
\begin{center}\begin{tabular}{c}\hline
#1 \\ \hline \end{tabular} \end{center}}

\begin{document}
\begin{titlepage}
\noindent
{\sl Physical Review Letters}
\hfill LBL-36358\\[8ex]
\begin{center}
{\large {\bf
Incorporation of Quantum Statistical Features \\
in Molecular Dynamics$^*$}}\\[8ex]
{\sl Akira Ohnishi}\\[1.5ex]
Department of Physics, Hokkaido University\\
Sapporo 060, Japan\\[2ex]
and\\[2ex]
{\sl J\o rgen Randrup}\\[1.5ex]
Nuclear Science Division, Lawrence Berkeley Laboratory\\
University of California, Berkeley, California 94720, USA\\[2ex]

\today\\[4ex]
{\sl Abstract:}
\\
\end{center}

{\small\noindent
We formulate a method for incorporating quantum fluctuations
into molecular-dynamics simulations of many-body systems,
such as those employed for energetic nuclear collision processes.
Based on Fermi's Golden Rule,
we allow spontaneous transitions to occur between the wave packets
which are not energy eigenstates.
The ensuing diffusive evolution in the space of the wave packet parameters
exhibits appealing physical properties,
including relaxation towards quantum-statistical equilibrium.

\vfill
\noindent
$^*$
This work was supported in part by the Director,
Office of Energy Research,
Office of High Energy and Nuclear Physics,
Nuclear Physics Division of the U.S. Department of Energy
under Contract No.\ DE-AC03-76SF00098,
the National Institute for Nuclear Theory
at the University of Washington in Seattle,
the Grant-in-Aid for Scientific Research (No.\ 06740193)
from Ministry of Education, Science and Culture, Japan,
and by Nukazawa Science Foundation.
The calculations were supported in part by RCNP, Osaka University,
as RCNP Computational Nuclear Physics Project No.\ 93-B-03.
}
\end{titlepage}

Molecular dynamics simulations are useful for understanding
both statistical and dynamical properties of many-body systems
in a variety of physical contexts \cite{Hoover}.
While quantitative insight can be obtained in many cases,
the foundation and interpretation of such approaches
are problematic when quantum systems are addressed.
In these approaches the many-body system is usually represented
as a (possibly antisymmetrized) product
of parametrized single-particle wave packets,
and equations of motion for the parameters are then derived
from a suitable variational principle.
This corresponds to a mean-field treatment of the quantal problem and
the ensuing parameter dynamics is then effectively classical.
Consequently,
the statistical properties of the system will be classical rather than quantal,
thus casting doubt on the quantitative utility of results
obtained in complicated scenarious where
quantal statistics plays a major role.

This generic shortcoming of molecular dynamics
originates in the neglect of the spectral distribution of energy eigenvalues
associated with the wave packets which are not energy eigenstates~\cite{OR93}.
In the present note we suggest a possible method
by which this inherent problem can be largely alleviated.
This novel method consists of introducing a stochastic term in the dynamics
so that a given wave packet may make spontaneous transitions
to neighboring wave packets in accordance with its spectral distribution,
and it is found that the ensuing diffusive evolution with this term
exhibits relaxation towards quantum-statistical equilibrium.
The method is rather general
and so it should be of correspondingly broad interest.

This issue is especially relevant in nuclear dynamics
where the system consists of nucleons at such densities and excitations
that quantum statistics plays a major role.
Indeed,
the interpretation of current heavy-ion collision experiments
depends on detailed dynamical simulations,
and so the problem is an urgent one.
In recent years,
significant effort has been devoted to the development
of microscopic simulation models for nuclear collisions,
of both one-body~\cite{guide} and $A$-body nature.
We shall address the situation in which a product of gaussian wave packets
are employed for the $A$-body system,
as has been done extensively in nuclear dynamics \cite{Aichelin,Boal,FMD,AMD},
but the proposed method is not restricted to this special case.

For notational convenience, we shall make our presentation within the
framework developed for
the Antisymmetrized Molecular Dynamics model~\cite{AMD},
and so the basic single-particle wave packets are gaussians of fixed width,
$<\r|\z>$,
where the real and imaginary parts of the parameter $\z$ specify
the centroid in position and momentum, respectively.
The normalized $A$-body product wave function,
$<\r_1,\cdots,\r_A|\Z>$,
is then characterized by the parameter vector $\Z=(\z_1,\cdots,\z_A)$.
The inclusion of antisymmetrization modifies the measure,
$d\Gamma={\rm det}(\C)d\Z$, where the matrix $\C$ has the elements
$C_{nn'}=\del^2\log{\cal N}/\del\zbar_n\del\z_{n'}$,
with $\cal N$ being the associated normalization constant;
the resolution of unity is then $\int d\Gamma |\Z><\Z|$.

With this convenient formalism,
the equations of motion for the wave packet parameters can then
be written on a compact form,
\beq\label{EoM}
i\hbar\C\cdot\dot{\Z}={\del{\cal H}\over\del\Zbar}\ ,
\eeq
where ${\cal H}=<\Z|\hat{H}|\Z>$ is the
expectation value of the $A$-body Hamiltonian operator $\hat{H}$ with
respect to the particular state $\Z$.
Though generally not of Hamiltonian form,
this system of equations produces a fully classical evolution.

The starting point for our present developments is the
quantum-mechanical feature that a given wave packet is generally not
an eigenstate of the many-body Hamiltonian.  The probability for the
wave packet $\Z$ to contain eigenstates of energy $E$ is given by the
spectral strength function,
\beq\label{rho}
\rhoE(\Z)\ \equiv\
<\Z|\delta(\hat{H}-E)|\Z>\ ,
\eeq
which is spread around the expectation value $\cal H$
with a variance given by
\beq\label{sigmaZ}
\sigma_Z^2=<\Z|(\hat{H}-{\cal H})^2|\Z> =\int d\Gamma'
|<\Z'|\hat{H}-{\cal H}|\Z>|^2 \approx {\del{\cal H}\over\del\Z} \cdot
\C^{-1} \cdot {\del{\cal H}\over\del\Zbar}\ .
\eeq

The equation of motion (\ref{EoM}) determines the
evolution of the wave packet parameter vector, $\Z(t)$, in an entirely
deterministic manner and without any physical effect of the spectral
structure of the wave packet.
In order to provide the system with an opportunity
for exploring and exploiting
the various eigencomponents
contributing to its wave packet,
we wish to augment the equation of motion by a stochastic term
that may cause occasional transitions between different wave packets.
Guided by Fermi's Golden Rule,
we then adopt the following form for the differential rate of transitions
from a given wave packet $\Z$ to others near $\Z'$,
\beq\label{rate}
w(\Z\rightarrow\Z')\ =\ {2\pi\over\hbar}\ |<\Z'|\hat{V}|\Z>|^2\ \rhoE(\Z')\ .
\eeq
Here the operator $\hat{V}$ represents a suitable ``residual'' interaction
and $E$ is a specified energy
which is usually taken as the expectation value
of the originally specified initial state.

When the above stochastic transitions are included in the dynamics,
the object of study is the distribution of
the wave packet parameter vector, $\phi(\Z,t)$.
For a closed (and sufficiently complex) system
this distribution will approach the associated equilibrium distribution.
Invoking the principle of detailed balance for a stationary distribution,
we readily see that the equilibrium distribution
is proportional to the spectral function $\rhoE(\Z)$.
Consequently,
the ensuing stochastic molecular dynamics populates the parameter space
in a microcanonical manner,
as is physically reasonable since the ensemble is characterized
by the specified energy $E$.
This feature is most easily recognized by considering
the microcanonical phase-space volume,
\beq
\Omega(E) \equiv {\rm Tr}\left(\delta(\hat{H}-E)\right)
=\int d\Gamma <\Z|\delta(\hat{H}-E)|\Z>\
=\int d\Gamma\ \rhoE(\Z)\ .
\eeq

For the discussion of statistical properties, it is
convenient to consider the associated canonical partition function
which is given by
\beq\label{Z}
{\cal Z}(\beta)\equiv \int_0^\infty dE\ \Omega(E)\ \rme^{-\beta E}
=\int_0^\infty dE\ \int d\Gamma\ \rhoE(\Z)\ \rme^{-\beta E}
=\int d\Gamma\ {\cal W}(\Z;\beta)\ .
\eeq
The the statistical weight of a given state
can thus be calculated once the form of the spectral density is known,
\beqar\label{W}
{\cal W}(\Z;\beta)\equiv\int_0^\infty dE\ \rhoE(\Z)\ \rme^{-\beta E}
\approx\exp\left[-{{\cal H}^2\over\sigma_Z^2}
(1-\rme^{-\beta\sigma_Z^2/{\cal H}})\right]\ .
\eeqar
The last relation holds exactly when the spectral strength distribution
is of Poisson form,
as is the case for a harmonic oscillator \cite{OR94}.

This latter result is very encouraging,
because the expression (\ref{W}) for the statistical weight
${\cal W}(\Z;\beta)$ leads to physically appealing statistical properties,
as already shown in ref.\ \cite{OR93}
and further discussed in ref.\ \cite{OR94}.
In order to illustrate this central point,
we show in fig.\ \ref{fig:1} the temperature dependence
of the mean excitation energy for a system of confined nucleons,
when a sampling of the wave packet space is performed
with the statistical weight (\ref{W}).
At low temperatures the system exhibits a typical quantal behavior,
with the energy rising as the square of the temperature.
As the temperature is increased the growth turns linear,
as is characteristic of classical systems.
This behavior should be contrasted with what would happen
without the spectral transitions,
\ie\ with the standard molecular dynamics.
Since the dynamics is then entirely classical,
the system will relax in accordance with the standard Boltzmann weight,
${\cal W}_{\rm class}(\Z;\beta)\sim\exp(-\beta{\cal H})$,
and its behavior would be classical throughout the entire temperature range
\cite{OR93}.
Thus,
the addition of the stochastic term (\ref{rate})
leads to dynamical evolutions that populate the parameter space
in better
accordance with quantum statistics.
We therefore expect that the incorporation of such stochastic transitions
into molecular-dynamics simulations may significantly improve
the description of features sensitive to the quantal fluctuations
in the many-body system,
such as the specific heat at low temperatures.

\figposition{Figure \ref{fig:1}}

In order to perform a practical implementation of the proposed
stochastic dynamics,
it is helpful to employ techniques from transport theory.
The introduction of the transitions governed by (\ref{rate})
leads to a diffusive transport process
in the space of the wave packet parameter vectors $\Z$.
The evolution of the associated distribution, $\phi(\Z,t)$,
can then be described by a Fokker-Planck equation,
\beq\label{FP}
{\del\over\del t}\phi(\z_1,\cdots,\z_A;t)=
-\sum_{n=1}^A {\del\over\del\z_n} V_n\phi\
+\ \sum_{nn'}^A{\del^2\over\del\z_n\del\zbar_{n'}}
D_{nn'}\phi\ .
\eeq
where the transport coefficients $V_n$ and $D_{nn'}$
can be calculated approximately as functions of $\Z$,
as we shall sketch below.

We first note that the residual interaction $\hat{V}$
in the expression (\ref{rate}) for the stochastic transition rate
should not have any diagonal matrix elements
(since such transitions would be spurious).
This can be accomplished by subtracting its expectation value
${\cal V}=<\Z|\hat{V}|\Z>$ before squaring.
It is then possible to show that the transition rate (\ref{rate})
can be written on the following convenient approximate form,
\beqar\label{w} \nonumber
w(\Z\rightarrow\Z+\delta\Z)&\approx&
{2\pi\over\hbar}
\left( {\del{\cal V}\over\del\Z}\cdot \delta\Z \right)
\left( \delta\Zbar \cdot {\del{\cal V}\over\del\Zbar} \right)
\rhoE(\Z)\\
\label{AppRate}
&\times& \exp\left[-\delta\Zbar\cdot\C\cdot\delta\Z
-\beta_Z(\delta\Zbar \cdot {\del{\cal H}\over\del\Zbar}
	+{\del{\cal H}\over\del\Z}\cdot \delta\Z )\right]\ ,
\eeqar
where $\beta_Z\equiv-\del\ln\rhoE/\del{\cal H}$
may be interpreted as a state-dependent temperature.

The total rate of transitions
from a given state $\Z$ into any other state $\Z'$
can then readily be calculated,
\beq\label{w0}
w_0(\Z)\ \equiv\ \int d\Gamma'\ w(\Z\rightarrow\Z')\
\approx\ {2\pi\over\hbar}\ \gamma_Z^2\
\rhoE(\Z)\ \rme^{\beta_Z^2\sigma_Z^2}\ ,
\eeq
where we have introduced the quantity
\beq\label{gammaZ}
\gamma_Z^2\ \equiv\ <\Z|(\hat{V}-{\cal V})^2|\Z>\
\approx\ {\del{\cal V}\over\del\Z}
\cdot \C^{-1} \cdot
{\del{\cal V}\over\del\Zbar}\ ,
\eeq
which can be regarded as a typical value
of the square of the transition matrix element in (\ref{rate}).
The expected number of transitions taking place during a small time interval
$\Delta t$ is then $n_0=w_0\Delta t$,
which may also be interpreted as the probability for any transition to occur
during $\Delta t$.

The transport coefficients entering in the Fokker-Planck equation (\ref{FP})
characterize the first and second moments
of the stochastic changes $\delta \z_n$
that have accumulated over the short time interval $\Delta t$,
when an average is taken over the entire ensemble of possible transitions
$\Z\rightarrow\Z'$,
\beqar
\prec\delta\z_n\succ\ &=& V_n(\Z)\ \Delta t\ ,\\
\prec\delta\z_n\delta\zbar_{n'}\succ\
&=& 2D_{nn'}(\Z)\ \Delta t\ .
\eeqar
Using the above simple expression (\ref{w}) for the basic transition rate,
we obtain the following results,
\beqar
\label{V}
V_n(Z) &\equiv&
	\int d\Gamma'\ \delta\z_n\ w\
	\approx \left( \bold{D}\cdot{\del\ln\rhoE\over\del\Zbar}\right)_n
	=-\beta_Z
	\left(\bold{D}\cdot{\del{\cal H}\over\del\Zbar}\right)_n\ ,\\
\nonumber
2D_{nn'}(\Z) &\equiv&
	\int d\Gamma'\ \delta\z_n \delta\zbar_{n'}\ w\
	\\
\label{D}
	&\approx& w_0
	\left[C^{-1}_{nn'} +
	{1\over\gamma_Z^2}
	\left(\bfC^{-1} \cdot {\del{\cal V}\over\del\Zbar}\right)_n
	\left( {\del{\cal V}\over\del\Z} \cdot \bfC^{-1}\right)_{n'} \right]\ .
\eeqar
The expression in the square bracket holds to the leading order
in $\beta_Z^2$.
It is easy to see that both the center-of-mass position
and the total momentum remain unchanged on the average,
$\sum_n V_n=0$,
whereas the individual histories will exhibit
diffusive Brownian-type excursions from the initial values,
due to the composite nature of the wave packets.
This behavior is to be expected,
since the the energy $\cal H$ is no longer a constant of motion
but will fluctuate around its initial value $E$.

The existence of the above approximate expressions
(\ref{w0}), (\ref{V}), and (\ref{D})
makes it a relatively easy task
to pick the stochastic changes $\delta\z_n$ at each time step
in the course of the dynamical evolution,
requiring only the diagonalization of the coefficient matrix $\C$.
Thus,
it is fairly easy to implement the proposed stochastic extension
and it may therefore be of practical utility.

Up to this point,
the presentation has been kept on a general level,
since the method is broadly applicable
and may be of interest in a variety of physical scenarios.
However,
since we were motivated by heavy-ion physics,
we wish to finally discuss how the proposed method may be of utility
in this particular subfield.
Generally,
the complexity of nuclear collisions necessitate microscopic simulations
for an informative interpretation of the data.
Currently,
considerable interest is focussed on socalled multifragmentation events,
in which the collision leads to the production of several
massive nuclear fragments.
It has proven difficult to reproduce this phenomenon
by ordinary molecular dynamics,
apparently because any massive fragments formed tend to be too excited
and, consequently, will quickly break up.
However,
if the presently proposed stochastic transitions are incorporated,
an excited massive fragment will explore its spectrum of eigenstates
and may thereby become trapped into more bound configurations,
thus leading to an enhanced survival probability.
In order to appreciate this mechanism,
it is important to recognize that the overall transition rate,
and the spectral spread of the transitions,
are proportional to the variance $\gamma_Z^2$
and so it generally increases with the
intrinsic excitation energy.
The chance for escaping from a well-bound configuration
is then smaller than the chance for deexciting into it,
as is consistent with detailed balance,
since the well-bound state has a higher statistical weight.
It thus appears very possible that the proposed model
may account better for the fragment yields.
We are presently exploring this central issue
by means of dynamical simulations \cite{next}.

In this note,
we have proposed a novel method for taking account of
the inherent energy spread associated with the wave packets
propagated in molecular-dynamics simulations of quantum many-body systems.
This simple physical idea is realized by augmenting
the standard deterministic equations of motion for the wave packet parameters
by a stochastic term that causes continual transitions between wave packets.
The resulting model is thus akin to the transport treatment of Brownian motion,
but it employs a Langevin force that originates in the quantal fluctuations
of the system.
The emerging dynamics exhibits appealing quantum-statistical features
and is therefore expected to present a significant advance
when complicated processes are addressed.
In particular,
application to nuclear multifragmentation processes
may yield dynamical evolutions
that are in qualitatively better agreement with the observations.

\medskip

This work was supported in part by the Director,
Office of Energy Research,
Office of High Energy and Nuclear Physics,
Nuclear Physics Division of the U.S. Department of Energy
under Contract No.\ DE-AC03-76SF00098,
the National Institute for Nuclear Theory
at the University of Washington in Seattle,
and the Grant-in-Aid for Scientific Research (No.\ 06740193)
from Ministry of Education, Science and Culture, Japan.
One of the authors (A.O.) also thanks Nukazawa Science Foundation
for its partial support for his visit to INT.
The calculations in this work were supported
by Research Center for Nuclear Physics (RCNP), Osaka University,
as RCNP Computational Nuclear Physics Project No.\ 93-B-03.

\newpage
\bfig[b]
\caption{Excitation energy versus temperature.}
\label{fig:1}
A system of 20 protons and 20 neutrons
is confined within a sphere of radius $40^{1/3}r_0$
and a Metropolis sampling is then performed
of the corresponding anti-symmetrized gaussian wave packets,
based on the modified statistical weight ${\cal W}(\Z;\beta)$
given in eq.\ (\ref{W}).
The abscissa is the imposed temperature $T=1/\beta$
and the ordinate is the calculated mean excitation energy
$\SEV{E}\equiv -\del \log \Part(\beta)/\del \beta$
(using the partition function (\ref{Z})
and with the ground-state energy subtracted),
and divided by the corresponding energy of a system of free nucleons,
$E_{\rm free}=40\times{3\over2}T$ (dashed line).
The solid line has been obtained with the nuclear Fermi-gas formula,
$E^*=aT^2$,
using the level density parameter $a=40/(8\ \MeV)$.
\efig

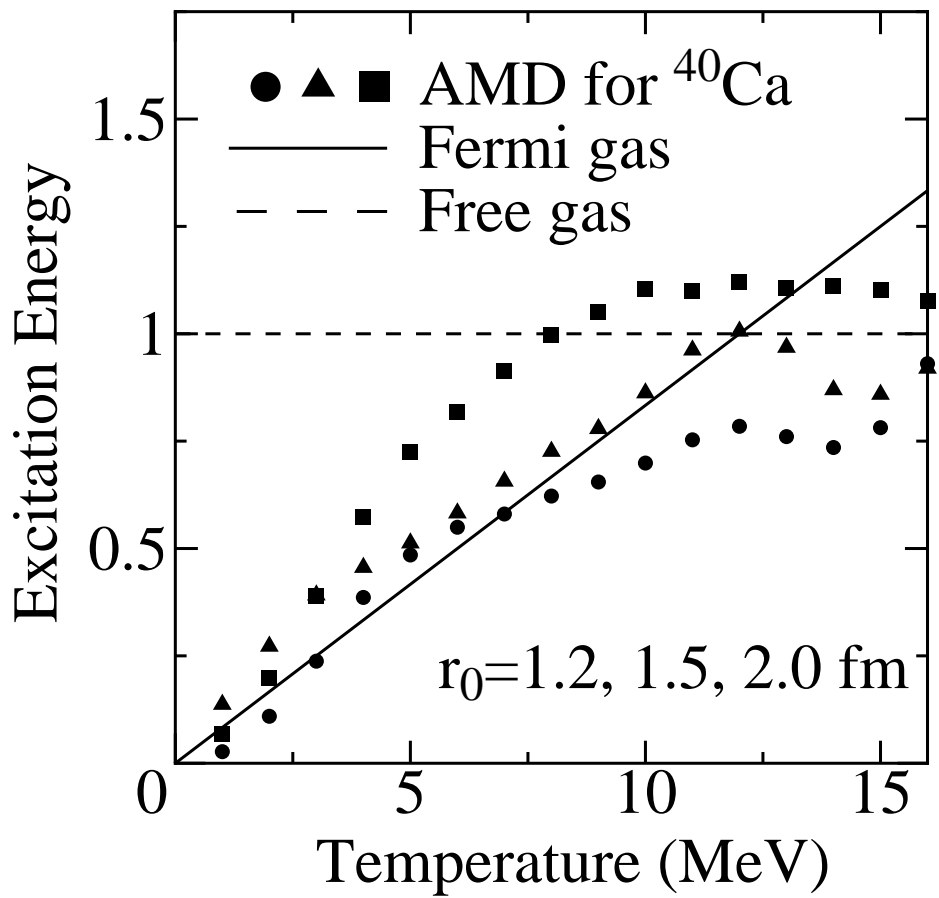
\end{document}